\documentclass[11pt,a4paper]{article}
\usepackage[margin=1in]{geometry}
\usepackage[utf8]{inputenc}
\usepackage[T1]{fontenc}
\usepackage{abstract}
\usepackage{amsmath,amssymb,amsfonts}
\usepackage{booktabs}
\usepackage{longtable}
\usepackage{graphicx}
\usepackage{adjustbox}
\usepackage{caption}
\usepackage{float}
\usepackage{rotating}
\usepackage{pdflscape}
\usepackage{appendix}
\usepackage[numbers,sort&compress]{natbib}
\bibpunct[, ]{[}{]}{,}{n}{,}{,}

\usepackage{siunitx}
\usepackage{array}
\usepackage{multirow}
\usepackage{hyperref}
\usepackage{bookmark}

\usepackage{setspace}
\title{External Demand, Domestic Monetary Conditions, and Remittance Dynamics in Nepal}
\author{Sahaj Raj Malla \\
        Department of Mathematics, Kathmandu University, \\
        P.O. Box 6250, Dhulikhel, Kavre, Bagmati Province, Nepal \\
        \texttt{sm03200822@student.ku.edu.np}}
\date{}

\begin{document}

\maketitle

\begin{abstract}
This study investigates the macroeconomic determinants and dynamic behaviour of personal remittances as a share of Gross Domestic Product (GDP) in Nepal, emphasizing external demand in major destination countries and domestic monetary policy. Using annual data (1993-2024), we construct composite indices via Principal Component Analysis (PCA) for multi-country external demand and a domestic Monetary Conditions Index (MCI). Our small-sample econometric pipeline includes Autoregressive Distributed Lag (ARDL) bounds testing, Engle-Granger cointegration, Dynamic OLS (DOLS), and a two-step Error Correction Model (ECM). We also employ Granger causality tests and multi-model forecasting using machine learning and ECM scenarios. The analysis reveals a strong positive long-run effect of external demand on remittances and a significant negative impact of tighter domestic monetary conditions. The ECM confirms a stable cointegrating relationship, correcting approximately 26\% of disequilibria annually. Medium-term projections indicate remittances will remain structurally important, reaching around 28.3\% of GDP by 2030 under baseline conditions, while exhibiting high sensitivity to external demand shocks. This study advances the literature by integrating PCA-derived external demand and monetary conditions indices within a unified ARDL-ECM framework for small samples. Focusing on one of the world's most remittance-dependent economies, it offers actionable insights for monetary policy calibration, migration diversification, and the productive utilization of remittance inflows.
\end{abstract}


\newpage

\section{Introduction}

Remittances constitute one of the most critical and stable sources of external financing for many developing economies, frequently surpassing foreign direct investment, official development assistance, and export revenues in both magnitude and resilience \citep{ratha2005, worldbank2023}. Nepal exemplifies this phenomenon to a remarkable degree. Over the past three decades, personal remittances have emerged as the cornerstone of the Nepalese economy, consistently accounting for more than 20--30 percent of Gross Domestic Product (GDP). These inflows have played a pivotal role in financing persistent trade deficits, supporting household consumption, reducing poverty, and bolstering macroeconomic stability amid structural challenges such as limited export competitiveness, political instability, and climate-related vulnerabilities.

Despite their central macroeconomic importance, the determinants of remittance flows to Nepal remain insufficiently understood within a robust multivariate time-series framework. Existing studies have largely emphasised micro-level motivations (altruism versus self-interest) or broad macroeconomic correlates such as income differentials, exchange rates, and political instability. While valuable, these approaches often overlook the joint dynamics between global external demand conditions (particularly in major migrant destination countries) and domestic monetary policy environments. The small-sample nature of annual macroeconomic data available for Nepal poses significant methodological challenges that many prior analyses have not fully addressed.

This study addresses these critical gaps by examining the \emph{determinants and dynamic behaviour} of personal remittances to Nepal, with particular emphasis on external demand conditions in migrant destination countries and the stance of domestic monetary policy. We construct theoretically grounded composite indices using Principal Component Analysis (PCA): a multi-country external demand index based on real GDP series from 12 major destination countries and a domestic Monetary Conditions Index (MCI) that captures the overall tightness of monetary policy. These indices effectively mitigate multicollinearity issues prevalent when using single-variable proxies.

The empirical analysis employs a comprehensive, publication-grade econometric pipeline tailored for small-sample environments. We implement rigourous stationarity and structural break testing, the Autoregressive Distributed Lag (ARDL) bounds testing approach of \cite{pesaran2001bounds}, supplemented by the Engle-Granger residual-based cointegration test, long-run estimation via Ordinary Least Squares (OLS) with Heteroskedasticity and Autocorrelation Consistent (HAC) standard errors, Dynamic OLS (DOLS), and a two-step Error Correction Model (ECM). The study also conducts extensive robustness checks across multiple specifications, performs Granger causality tests, and develops a multi-model forecasting exercise, integrating univariate time-series models, machine learning ensembles, and scenario-based ECM projections to 2030.

This study makes several important contributions to the literature:
\begin{itemize}
    \item This study systematically integrates PCA-derived composite indices of external demand and domestic monetary conditions within a unified ARDL-ECM framework for Nepal.
    \item It applies a comprehensive suite of modern time-series techniques suitable for small samples while explicitly addressing data limitations through imputation and structural break handling.
    \item It extends the analysis to medium-term forecasting under alternative global and domestic scenarios, thereby bridging academic rigour with forward-looking policy relevance in one of the world’s most remittance-dependent economies.
\end{itemize}

The remainder of the study is organised as follows. Section \ref{sec:literature} reviews the relevant theoretical and empirical literature on remittance determinants. Section \ref{sec:methodology} describes the data sources, variable construction, and econometric methodology. Section \ref{sec:results} presents the empirical results, including diagnostics, cointegration tests, long-run and short-run estimates, and robustness checks. Section \ref{sec:forecasting} discusses the multi-model forecasting framework and projections to 2030. Section \ref{sec:discussion} interprets the findings in light of theory and draws policy implications, while Section \ref{sec:conclusion} concludes.


\section{Literature Review}
\label{sec:literature}

Remittances represent a vital and relatively stable source of external financing for developing economies, often exhibiting greater stability and counter-cyclical properties compared to other international capital flows \citep{ratha2005, worldbank2023}. The theoretical foundations of remittance behaviour draw from several complementary frameworks. The \textit{altruism} model posits that migrants send money home primarily to support family welfare, with remittances responding positively to adverse economic conditions in the home country \citep{lucas1985}. In contrast, the \textit{self-interest} or investment motive emphasises portfolio considerations, including expected returns on home-country assets, exchange rate movements, and relative economic performance between host and home countries \citep{stark1985, amuedo2006}. A more integrated \textit{portfolio diversification} perspective views remittances as part of a broader risk-sharing and consumption-smoothing strategy within transnational households \citep{rapoport2006}.

\subsection{Empirical Literature on Remittance Determinants}

Cross-country empirical studies have consistently identified host-country economic conditions, particularly income levels, labour market performance, and commodity prices, as key ``pull'' factors driving remittance flows \citep{singh2011, mcgowan2010determinants, combes2012}. On the home-country side, factors such as exchange rate depreciation, financial development, and political stability have been shown to influence both the volume and stability of remittance inflows \citep{aggarwal2011, nyasha2022impact}.

A growing body of research highlights the importance of global and regional demand conditions. Strong economic performance in destination countries increases migrant employment opportunities and earnings, thereby boosting remittance outflows \citep{mcgowan2010determinants}. Conversely, recessions or tighter immigration policies in host countries can lead to significant contractions in flows. Domestic monetary conditions in recipient countries, however, remain relatively understudied. Tighter monetary policy (higher interest rates or reserve requirements) may raise the opportunity cost of holding domestic assets or constrain credit availability, potentially affecting household decisions regarding remittance receipt and utilisation \citep{aggarwal2011}.

\subsection{Nepal-Specific Literature}

Nepal provides a compelling case study due to its exceptionally high dependence on remittances, which have averaged over 25\% of GDP in recent years. Early studies documented the poverty-reducing and growth-enhancing effects of remittances at both household and national levels \citep{lokshin2010}. Subsequent research has examined the impact of remittances on financial development, exchange rate dynamics, and potential Dutch disease effects \citep{sapkota2013}.

Few studies have systematically incorporated composite indices of external demand or rigourously examined the role of domestic monetary conditions. While several studies acknowledge the importance of GCC economies and oil prices, rigourous multi-country external demand proxies constructed using dimensionality reduction techniques such as PCA are notably absent. Similarly, the interaction between Nepal’s domestic monetary policy stance and remittance inflows has received limited empirical scrutiny.

\subsection{Research Gaps and Contributions of This Study}

This study addresses several critical gaps in the existing literature. 
\begin{itemize}
    \item It develops theoretical composite indices using PCA for both multi-country external demand (across 12 major destination countries) and domestic monetary conditions, thereby mitigating multicollinearity issues that commonly plague single-variable proxy approaches.
    \item It implements a comprehensive small-sample econometric pipeline, including stationarity testing with structural break consideration, ARDL bounds testing \citep{pesaran2001bounds}, DOLS, a two-step ECM, extensive robustness checks, and multi-model forecasting, specifically tailored to the constraints of Nepalese annual macroeconomic data.
    \item By combining rigourous cointegration analysis with forward-looking scenario-based projections to 2030, the study bridges the gap between academic rigour and policy relevance in one of the world’s most remittance-dependent economies.
\end{itemize}

By integrating external demand conditions and domestic monetary policy within a unified cointegration and error-correction framework, this study offers a more complete understanding of the structural drivers and dynamic adjustment mechanisms of remittances to Nepal. The findings are expected to contribute meaningfully to both the academic literature on migration and development economics and to evidence-based policymaking in Nepal.


\section{Data and Methodology}
\label{sec:methodology}

This study utilises annual time-series data for Nepal covering the period 1993--2024, yielding 32 observations. The dataset integrates domestic macroeconomic indicators, external demand conditions from major migrant destination countries, global commodity prices, and detailed monetary policy variables. All series are harmonised to calendar years, with fiscal-year data aligned to the ending calendar year, consistent with standard practice in Nepalese macroeconomic research.

\subsection{Data Sources and Variables}

The primary dependent variable is personal remittances received as a percentage of GDP, sourced from the World Bank World Development Indicators (WDI). This measure captures formal remittance inflows and is the standard variable employed in the international remittances literature.

Domestic macroeconomic controls include real GDP growth (\%), Consumer Price Index (CPI) inflation rate (\%), unemployment rate (modelled ILO estimate), total population, and the official exchange rate (NPR per USD, period average). Monetary policy variables, obtained from Nepal Rastra Bank publications and statistical bulletins, comprise the policy (Bank) rate, interbank rate, 91-day Treasury bill rate, weighted average deposit and lending rates, interest rate spread, and Cash Reserve Ratio (CRR) for commercial banks.

External demand conditions are proxied through real GDP (constant US\$) series for 12 major migrant destination countries: Qatar, India, United Arab Emirates, Saudi Arabia, Malaysia, United States, Japan, Kuwait, Bahrain, South Korea, United Kingdom, and Australia, all sourced from the World Bank WDI. Brent crude oil price (USD per barrel, annual average), obtained from the U.S. Energy Information Administration (EIA), serves as a key global commodity indicator influencing Gulf labour markets.

\subsection{Imputation and Transformations}

Several preprocessing steps were implemented to ensure data quality. Missing observations in the monetary policy block for the early period (1993--2000) were imputed using a combination of official Nepal Rastra Bank historical reference rates and backward linear extrapolation based on observed trends from 2001--2005. All imputations were transparently documented and subjected to sensitivity analysis in robustness checks.

To stabilise variance, reduce heteroskedasticity, and facilitate elasticity interpretation, natural logarithmic transformations were applied to remittances (\% of GDP), population, oil price, exchange rate, and the GDP series of all destination countries. The natural logarithm of remittances as a percentage of GDP serves as the dependent variable in the core econometric specifications.

\subsection{Construction of Composite Indices}

Given the high degree of multicollinearity among external and domestic variables, PCA was employed to construct theoretically grounded composite indices.

\textbf{External Demand Indices:} A comprehensive external demand index was derived from the first principal component (PC1) of the log-transformed real GDP series of the 12 destination countries. This component explains a substantial proportion of the total variance and serves as the primary proxy for global pull factors. For robustness, a Gulf-specific sub-index (Qatar, UAE, Saudi Arabia, Kuwait, Bahrain, and Malaysia) and a trade-weighted external demand index (base year 2010 = 1.0, using destination-specific migrant remittance shares as weights) were also constructed.

\textbf{Monetary Conditions Index (MCI):} A domestic Monetary Conditions Index was constructed by applying PCA to five key financial indicators: the policy rate, interbank rate, 91-day Treasury bill rate, deposit rate, and lending rate. PC1, which captures the overall stance of monetary policy (with higher values indicating tighter conditions), explains approximately 64\% of the variance and is retained as the main monetary policy variable.

These composite indices effectively reduce dimensionality while preserving the core economic signals, thereby enhancing model stability and interpretability in the presence of multicollinear regressors.

\subsection{Econometric Strategy}

The empirical analysis follows a systematic, multi-step econometric pipeline specifically designed for small-sample time-series data:

First, stationarity properties of all variables were examined using the Augmented Dickey-Fuller (ADF), Phillips-Perron (PP), Kwiatkowski-Phillips-Schmidt-Shin (KPSS), and Zivot-Andrews structural break-aware unit root tests. Structural breaks were formally identified through sequential Chow tests at economically significant years (2002, 2008, 2015, and 2020), with a dominant break detected in 2002. This break is controlled for using impulse and step dummies.

Second, the presence of a long-run cointegrating relationship was tested using the ARDL bounds testing approach proposed by \cite{pesaran2001bounds} (Case III, maximum lags $p=q=2$), supplemented by the Engle-Granger residual-based cointegration test.

Upon confirmation of cointegration, long-run parameters were estimated using OLS with HAC standard errors. Diagnostic statistics including, Variance Inflation Factors (VIF), joint significance Wald tests, and bootstrap confidence intervals (1,000 replications), were computed. DOLS with leads and lags was employed as a robustness check against potential endogeneity and small-sample bias \citep{stock1993}.

Short-run dynamics and adjustment mechanisms were modelled through a two-step ECM, incorporating the lagged error correction term derived from the long-run residuals, short-run differenced variables, and structural break dummies. The speed of adjustment and half-life of deviations from long-run equilibrium were calculated.

Model reliability was assessed through a comprehensive battery of residual diagnostic tests: Breusch-Godfrey LM test for serial correlation, ARCH LM and Breusch-Pagan tests for heteroskedasticity, Jarque-Bera test for normality, and Ramsey RESET test for functional form. Parameter stability was evaluated using the CUSUM and CUSUMSQ tests. Robustness was examined across seven alternative specifications, including alternative demand proxies, subsample analyses, exclusion of major shock years, and level-dependent variable estimation.

Directionality of causal relationships was examined through pairwise Granger causality tests (maximum lag of 2). Finally, medium-term forecasts to 2030 were generated using a multi-model framework comprising univariate time-series models (ARIMA, ETS, Theta), machine learning ensembles (Ridge, Lasso, ElasticNet, Gradient Boosting), and scenario-based projections from the estimated ECM under Baseline, High External Demand, and Low External Demand scenarios.

All estimations and visualisations were implemented in Python 3.13 using \texttt{statsmodels}, \texttt{scikit-learn}, \texttt{pandas}, and \texttt{matplotlib}. This rigourous methodological approach ensures robust inference despite the modest sample size inherent to annual macroeconomic data for Nepal.


\section{Empirical Results}
\label{sec:results}

This section presents the core empirical findings of the study. Results are reported sequentially, covering descriptive statistics and correlations, stationarity and structural break analysis, cointegration tests, long-run determinants, short-run dynamics, diagnostic validation, robustness checks, and Granger causality.

\subsection{Descriptive Statistics and Correlations}

Table \ref{tab:descriptive} summarises the key variables over the period 1993--2024. Remittances as a percentage of GDP averaged 15.77\%, ranging from under 1\% in the mid-1990s to over 27\% in recent years, reflecting the growing structural importance of migration for the Nepalese economy. The external demand PCA index and Gulf-focused index exhibit very strong positive correlations with remittances ($\rho \approx 0.918$ and $0.920$). The official exchange rate, measured as the Nepalese rupee per U.S. dollar (NPR/USD), and the U.S. dollar (USD) series are incorporated as key macroeconomic variables.

\begin{table}[!htbp]
\centering
\caption{Descriptive Statistics of Key Variables}
\label{tab:descriptive}
\begin{tabular}{lrrrrr}
\toprule
Variable & N & Mean & Std. Dev. & Min & Max \\
\midrule
Remittances (\% of GDP) & 32 & 15.7747 & 10.0496 & 0.9767 & 27.6261 \\
ln(Remittances \% GDP) & 32 & 2.2788 & 1.2485 & -0.0236 & 3.3188 \\
External Demand Index (PCA-12) & 32 & 0.0000 & 3.3239 & -5.8296 & 4.2724 \\
External Demand (Gulf PCA-6) & 32 & 0.0000 & 2.4721 & -4.0524 & 3.0332 \\
External Demand (Trade-Weighted) & 32 & 0.8998 & 0.5161 & 0.2437 & 1.7786 \\
ln(Oil Price, USD/bbl) & 32 & 3.8259 & 0.6796 & 2.5463 & 4.7152 \\
ln(NPR/USD) & 32 & 4.3855 & 0.2930 & 3.8210 & 4.8913 \\
Monetary Conditions Index & 32 & 0.0000 & 1.8131 & -2.3169 & 4.2348 \\
Inflation Rate (\%) & 32 & 6.6619 & 2.6538 & 2.2700 & 11.2400 \\
Real GDP Growth (\%) & 32 & 4.3131 & 2.2526 & -2.3700 & 8.9800 \\
Unemployment Rate (\%) & 32 & 10.7022 & 0.5233 & 10.4370 & 13.0370 \\
\bottomrule
\end{tabular}
\end{table}

\subsection{Unit Root and Structural Break Analysis}

Unit root tests confirm that the dependent variable and most core regressors are integrated of order one, I(1), while inflation is I(0). No variable is I(2), validating the use of the ARDL bounds testing procedure. Sequential Chow tests identify a highly significant structural break in 2002 ($F=77.11$, $p<0.001$), consistent with major shifts in Nepal’s labour migration policy. Additional breaks were detected around 2008. These breaks are explicitly controlled for using impulse and step dummies in subsequent estimations.

\subsection{Cointegration Tests}

The ARDL bounds test (Case III, $p=2$, $q=2$) yields an F-statistic of 6.0658, which exceeds the 1\% upper-bound critical value of 4.43. We therefore reject the null hypothesis of no cointegration at the 1\% significance level. This finding is corroborated by the Engle-Granger residual-based test ($\tau = -3.4502$, $p \approx 0.0006$), confirming the existence of a stable long-run equilibrium relationship among remittances, external demand, and monetary conditions.

\subsection{Long-Run Determinants}

Table \ref{tab:longrun} presents the long-run estimates obtained from OLS with HAC standard errors. The external demand PCA index exerts a positive and statistically significant effect ($\beta = 0.246$, $p=0.035$), indicating that a one-standard-deviation increase in destination-country economic activity raises remittances as a share of GDP by approximately 0.25 percentage points in the long run. Tighter domestic monetary conditions have a strong negative impact ($\beta = -0.211$, $p<0.001$). The joint significance of the regressors is confirmed by a Wald test ($F=35.32$, $p<0.001$).

\begin{table}[!htbp]
\centering
\caption{Long-Run Estimates (HAC Standard Errors)}
\label{tab:longrun}
\begin{tabular}{lrrrr}
\toprule
Variable & Coefficient & HAC SE & t-statistic & p-value \\
\midrule
External Demand (PCA-12) & 0.2460 & 0.1169 & 2.10 & 0.035 \\
ln(Oil Price) & 0.0968 & 0.2871 & 0.34 & 0.736 \\
ln(NPR/USD) & 1.1112 & 0.9047 & 1.23 & 0.219 \\
Monetary Conditions Index & -0.2109 & 0.0520 & -4.06 & 0.000 \\
Inflation & -0.0001 & 0.0150 & -0.00 & 0.996 \\
\bottomrule
\end{tabular}
\end{table}

DOLS estimation reinforces these conclusions with greater precision. The coefficient on external demand increases to 0.316 ($p<0.001$) and the monetary conditions coefficient strengthens to -0.284 ($p<0.001$), while the exchange rate becomes marginally significant.

\subsection{Short-Run Dynamics (Error Correction Model)}

The two-step Error Correction Model exhibits excellent fit ($R^2=0.8975$). The error correction term is negative and highly significant ($\psi = -0.263$, $p=0.0016$), implying that approximately 26.3\% of any disequilibrium is corrected within one year (half-life $\approx$ 2.28 years). The 2002 impulse dummy is statistically important and helps stabilise short-run dynamics.

\subsection{Diagnostic Tests and Robustness Checks}

Residual diagnostics are generally satisfactory. There is no evidence of serial correlation (Breusch-Godfrey lag-2 $p=0.378$) or conditional heteroskedasticity (ARCH LM $p=0.656$). The model passes the Ramsey RESET test ($p=0.763$). Although normality is rejected (Jarque-Bera $p=0.019$), this is common in macroeconomic time series and does not undermine consistency. CUSUM and CUSUMSQ tests (see Appendix) indicate parameter stability.

Robustness checks across seven alternative specifications, including Gulf-focused and trade-weighted demand proxies, replacement of the MCI with individual policy variables, exclusion of major shock years, post-2000 subsample, and a level-dependent variable specification, which confirms the stability of the core results. The error correction term remains negative and statistically significant in all models. Granger causality tests support unidirectional causality running from external demand and monetary conditions to remittances. The empirical evidence collectively demonstrates that external demand conditions and domestic monetary policy are the dominant drivers of remittance inflows to Nepal.


\section{Forecasting and Scenario Analysis}
\label{sec:forecasting}

A key contribution of this study is the development of a multi-model forecasting framework to project Nepal’s remittance inflows as a percentage of GDP through 2030. This forward-looking analysis combines univariate time-series models, machine learning ensembles, and structurally grounded ECM scenarios, providing policymakers with a comprehensive range of plausible trajectories.

\subsection{Univariate Time-Series Models}

Out-of-sample predictive performance was evaluated on the final five observations (2020--2024). Among the univariate models, Exponential Smoothing (ETS) with damped trend outperformed both ARIMA and the Theta method, achieving the lowest Mean Absolute Percentage Error (MAPE = 6.385\%) and Root Mean Square Error (RMSE = 0.216). This model was selected as the best univariate benchmark.

\subsection{Machine Learning Ensemble}

Four machine learning models (Ridge, Lasso, ElasticNet, and Gradient Boosting Regressor) were trained on lagged features of remittances, external demand, oil prices, exchange rate, and monetary conditions. Ridge Regression demonstrated the strongest out-of-sample performance with an RMSE of 0.200, indicating good generalisation despite the modest sample size.

\subsection{ECM Scenario Projections (2025--2030)}

Leveraging the estimated Error Correction Model, scenario-based forecasts were generated under three plausible macroeconomic environments:

\begin{itemize}
    \item \textbf{Baseline Scenario}: Exogenous variables evolve according to their historical average growth rates.
    \item \textbf{High External Demand Scenario}: Stronger economic performance in destination countries (+1 standard deviation shock to the external demand index) combined with moderately lower oil prices.
    \item \textbf{Low External Demand Scenario}: Global economic slowdown (-1 standard deviation shock to external demand) accompanied by tighter domestic monetary conditions.
\end{itemize}

Table \ref{tab:forecast} presents the consolidated projections. Under the baseline ECM scenario, remittances as a percentage of GDP are expected to reach approximately \textbf{28.27\% by 2030}. The High External Demand scenario raises this figure to 29.43\%, while the Low External Demand scenario constrains it to 25.62\%.

\begin{table}[!htbp]
\centering
\caption{Forecast Comparison: Remittances as \% of GDP (2025--2030)}
\label{tab:forecast}
\begin{tabular}{lrrrr}
\toprule
Year & Univariate (ETS) & ECM Baseline & ECM High Demand & ECM Low Demand \\
\midrule
2025 & 25.95 & 26.28 & 26.46 & 26.01 \\
2026 & 27.86 & 26.48 & 26.87 & 25.83 \\
2027 & 29.85 & 26.81 & 27.39 & 25.72 \\
2028 & 31.93 & 27.24 & 28.01 & 25.66 \\
2029 & 34.09 & 27.73 & 28.69 & 25.63 \\
2030 & 36.33 & 28.27 & 29.43 & 25.62 \\
\bottomrule
\end{tabular}
\end{table}

\begin{figure}[!htbp]
\centering
\includegraphics[width=0.92\textwidth]{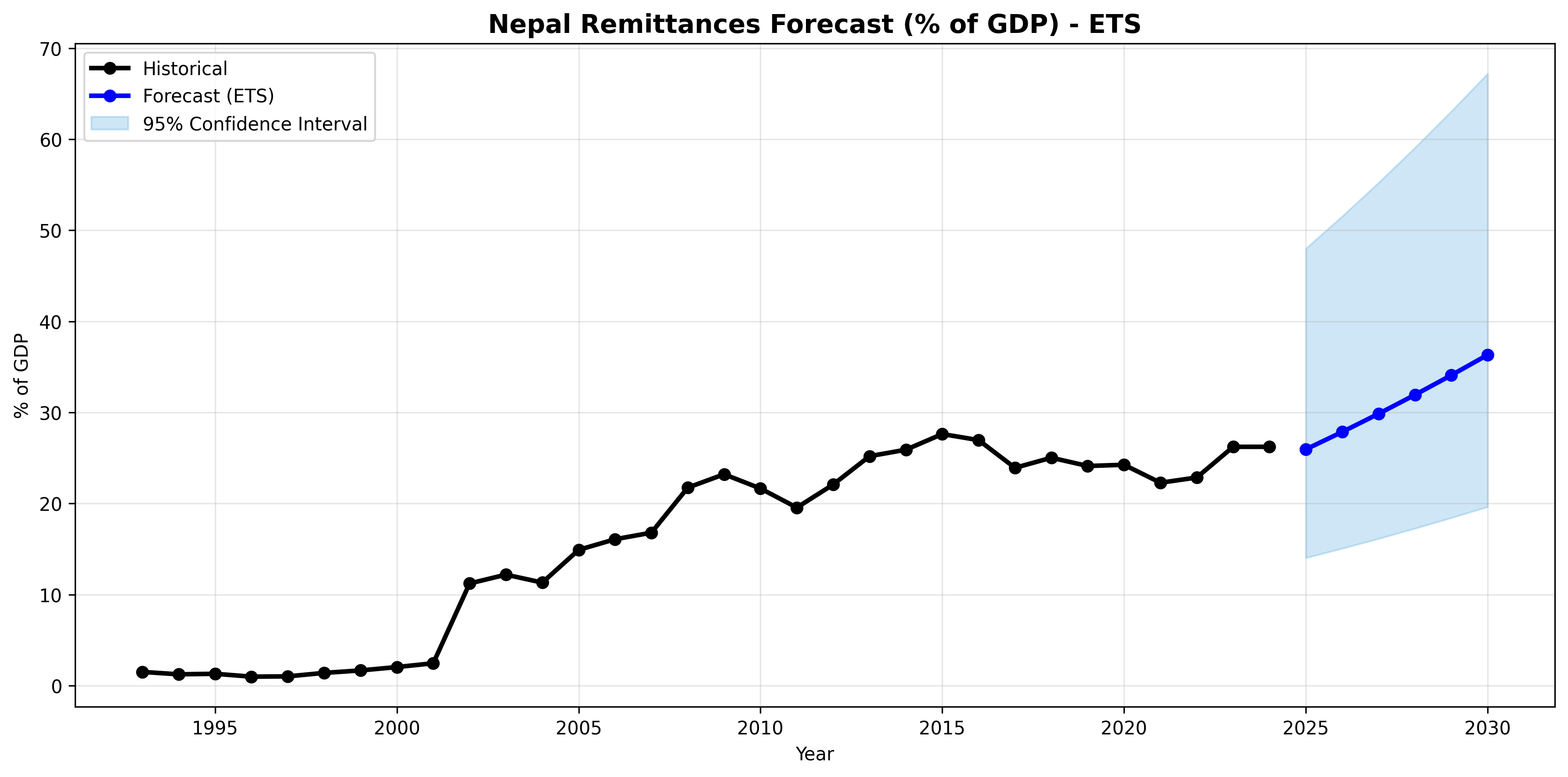}
\caption{Historical Remittances (\% of GDP) and Multi-Model Forecasts to 2030}
\label{fig:forecast}
\end{figure}

The divergence between scenarios highlights Nepal’s structural vulnerability to global economic cycles in migrant destination countries. While baseline projections suggest continued stability, adverse external shocks could significantly dampen future inflows. The multi-model framework also serves as an indirect validation of the structural ECM, demonstrating reasonable out-of-sample behaviour.

The forecasting analysis reinforces the central empirical findings: external demand conditions and domestic monetary policy are the primary drivers of remittance dynamics in Nepal, with important implications for medium-term macroeconomic planning, foreign exchange reserve management, and migration policy.

\section{Discussion}
\label{sec:discussion}

The empirical results of this study provide robust evidence regarding the determinants and dynamic behaviour of personal remittances to Nepal. External demand conditions in major migrant destination countries and domestic monetary policy emerge as the primary structural drivers of remittance inflows. These findings carry important theoretical, empirical, and policy implications.

\subsection{Interpretation of Main Findings}

The strong positive long-run coefficient on the external demand PCA index confirms that economic performance in destination countries acts as a powerful pull factor for remittance flows. A one-standard-deviation increase in the external demand index is associated with an increase in remittances as a share of GDP by approximately 0.25 percentage points (and 0.32 percentage points according to DOLS estimates). This result aligns with the new economics of labour migration theory \citep{stark1985, rapoport2006}, whereby higher migrant earnings and employment opportunities abroad directly translate into larger remittance transfers. The robustness of this relationship across multiple demand proxies (Gulf-specific PCA and trade-weighted index) underscores its centrality in Nepal’s remittance dynamics.

Conversely, the consistently negative and highly significant coefficient on the Monetary Conditions Index highlights the friction role of domestic monetary policy. Tighter monetary conditions appear to discourage or constrain remittance inflows, possibly through increased opportunity costs of holding domestic assets, reduced credit availability for remittance-receiving households, or altered migrant return intentions. This finding extends the existing literature by demonstrating that domestic monetary policy is not neutral with respect to remittance behaviour in highly remittance-dependent economies.

The Error Correction Model reveals a stable long-run cointegrating relationship with rapid adjustment: approximately 26.3\% of any disequilibrium is corrected within one year, with a half-life of deviations from equilibrium of roughly 2.28 years. The significant structural break in 2002, successfully captured through impulse dummies, emphasises the importance of accounting for major policy shifts in migration regimes when analysing Nepalese macroeconomic time series.

\subsection{Consistency with Existing Literature}

These results are broadly consistent with global evidence on remittance determinants \citep{mcgowan2010determinants, singh2011, combes2012}, which emphasises host-country economic conditions as key drivers. They also reinforce Nepal-specific findings on the importance of external factors while offering new insights into the understudied role of domestic monetary conditions \citep{aggarwal2011}.

\subsection{Policy Implications}

The findings have several actionable implications for Nepalese policymakers:

\begin{itemize}
    \item Given the strong dependence on external demand, efforts to diversify migrant destinations beyond the Gulf Cooperation Council countries and Malaysia should be prioritised. Negotiating new bilateral labour agreements, investing in skills development aligned with emerging global labour market needs, and promoting safe, orderly, and regular migration can reduce vulnerability to regional economic cycles.
    \item The negative impact of tighter monetary conditions suggests that Nepal Rastra Bank should carefully consider the indirect effects of policy rate hikes and liquidity tightening on remittance inflows and household welfare. A balanced monetary policy approach that maintains price stability while supporting financial deepening and credit access for remittance-dependent households appears warranted.
    \item The projected continuation of high remittance dependence (around 28.3\% of GDP by 2030 under baseline conditions) calls for strategic utilisation of these inflows. Policies that promote financial inclusion, diaspora investment vehicles, and the channelling of remittances toward productive investment rather than consumption could enhance their contribution to long-term economic growth and macroeconomic resilience.
    \item The high sensitivity of forecasts to external shocks underscores the need for robust foreign exchange reserve management, counter-cyclical fiscal frameworks, and contingency planning.
\end{itemize}

\subsection{Limitations}

Some limitations of the present study should be acknowledged:

\begin{itemize}
    \item The analysis is based on annual data with a relatively small sample of 32 observations, which limits degrees of freedom and may constrain the statistical power of the estimations.
    \item Official remittance statistics capture only formal transfer channels and therefore exclude informal flows, which are likely to remain economically significant in the Nepalese context.
    \item Although PCA-based composite indices help mitigate multicollinearity and improve model parsimony, they may obscure country-specific heterogeneity in external demand conditions.
\end{itemize}

\subsection{Directions for Future Research}

Future research may usefully extend the present analysis in several directions:

\begin{itemize}
    \item Higher-frequency data, such as quarterly observations, could be employed where available to capture short-run dynamics more precisely.
    \item Micro-level household survey data could be incorporated to better understand remittance behaviour, usage patterns, and distributional effects at the household level.
    \item Non-linearities and threshold effects, including regime-dependent responses to oil prices or external demand shocks, could be examined in greater depth.
    \item Comparative analysis with other highly remittance-dependent economies in South Asia and beyond could help place the Nepalese case in a broader regional and international perspective.
\end{itemize}


\section{Conclusion}
\label{sec:conclusion}

This study has examined the determinants and dynamic behaviour of personal remittances to Nepal over the period 1993--2024 using a comprehensive small-sample multivariate time-series framework. By constructing PCA-based composite indices of external demand across 12 major migrant destination countries and domestic monetary conditions, and applying a rigourous econometric pipeline encompassing unit root and structural break testing, ARDL bounds testing, DOLS, a two-step Error Correction Model, extensive robustness checks, and multi-model forecasting, the analysis yields several robust findings. External demand conditions exert a strong positive long-run effect on remittances as a share of GDP, while tighter domestic monetary conditions significantly dampen inflows. The Error Correction Model confirms a stable cointegrating relationship with rapid adjustment, correcting approximately 26\% of disequilibria within one year. Medium-term projections indicate that remittances will remain structurally vital, reaching around 28.3\% of GDP by 2030 under baseline conditions, although highly sensitive to global economic cycles in destination countries. 

Methodologically, this study advances the literature by integrating theoretically grounded composite indices within a unified ARDL-ECM framework tailored for small samples. From a policy perspective, the findings underscore the need for migration destination diversification, careful calibration of monetary policy that accounts for its indirect effects on remittance inflows, and strategic utilisation of remittances to enhance macroeconomic resilience and sustainable development in one of the world’s most remittance-dependent economies. Future research could extend this work through higher-frequency data, non-linear specifications, and comparative analysis across South Asian countries.


\section*{Acknowledgements}
The author gratefully acknowledges Nepal Rastra Bank and the World Bank for providing access to data. The author also extends sincere gratitude to his mother, Basuna Malla, and his family for their unwavering emotional support throughout the research process.

\newpage

\bibliographystyle{plainnat}
\bibliography{references} 

\newpage


\appendix

\section{Additional Tables and Figures}

This appendix presents supplementary materials supporting the empirical analysis in the main text.

\subsection{Unit Root Test Results}

Detailed unit root test results are reported in Table \ref{tab:unitroot_app}.

\begin{table}[H]
\centering
\caption{Detailed Unit Root Test Results}
\label{tab:unitroot_app}
\resizebox{\textwidth}{!}{%
\begin{tabular}{lcccccc}
\toprule
Variable & ADF Level (p) & PP Level (p) & KPSS Level (p) & ADF Diff (p) & ZA Break Year & Integration Order \\
\midrule
ln(Remittances \% GDP) & 0.610 & 0.603 & 0.017 & 0.079 & -- & I(1) \\
External Demand (PCA-12) & 0.904 & 0.903 & 0.024 & 0.000 & 2010 & I(1) \\
ln(Oil Price) & 0.696 & 0.696 & 0.027 & 0.000 & 2021 & I(1) \\
ln(NPR/USD) & 0.598 & 0.598 & 0.100 & 0.000 & 2010 & I(1) \\
Monetary Conditions Index & 0.994 & 0.932 & 0.010 & 0.011 & 2023 & I(1) \\
Inflation Rate (\%) & 0.000 & 0.000 & 0.100 & 0.001 & -- & I(0) \\
\bottomrule
\end{tabular}%
}
\end{table}

\subsection{Long-Run Estimation (Full Results with Constant)}

\begin{table}[H]
\centering
\caption{Full Long-Run Estimation Results (HAC Standard Errors)}
\label{tab:longrun_full}
\resizebox{\textwidth}{!}{%
\begin{tabular}{lrrrrr}
\toprule
Variable & Coefficient & HAC SE & t-statistic & p-value & Significance \\
\midrule
Constant & -2.9643 & 4.7347 & -0.63 & 0.531 & \\
External Demand (PCA-12) & 0.2460 & 0.1169 & 2.10 & 0.035 & ** \\
ln(Oil Price) & 0.0968 & 0.2871 & 0.34 & 0.736 & \\
ln(NPR/USD) & 1.1112 & 0.9047 & 1.23 & 0.219 & \\
Monetary Conditions Index & -0.2109 & 0.0520 & -4.06 & 0.000 & *** \\
Inflation & -0.0001 & 0.0150 & -0.00 & 0.996 & \\
\bottomrule
\end{tabular}%
}
\end{table}

\subsection{Error Correction Model (Full Short-Run Dynamics)}

\begin{table}[H]
\centering
\caption{Error Correction Model -- Short-Run Dynamics}
\label{tab:ecm_full}
\begin{tabular}{lrrrr}
\toprule
Variable & Coefficient & HC1 SE & t-statistic & p-value \\
\midrule
Constant & 0.0234 & 0.0453 & 0.52 & 0.605 \\
ECT$_{t-1}$ & -0.2627 & 0.0831 & -3.16 & 0.002 \\
$\Delta$ ln(Remittances \% GDP)$_{t-1}$ & 0.1351 & 0.0835 & 1.62 & 0.106 \\
$\Delta$ External Demand (PCA-12) & 0.1081 & 0.1188 & 0.91 & 0.363 \\
$\Delta$ ln(Oil Price) & -0.1846 & 0.1586 & -1.16 & 0.244 \\
$\Delta$ ln(NPR/USD) & -0.3043 & 0.7590 & -0.40 & 0.688 \\
$\Delta$ Monetary Conditions Index & -0.0247 & 0.0377 & -0.65 & 0.513 \\
$\Delta$ Inflation & 0.0127 & 0.0117 & 1.09 & 0.275 \\
$D_{2002}$ (Impulse) & 1.3042 & 0.0552 & 23.61 & 0.000 \\
$D_{COVID}$ (2020) & -0.0130 & 0.0618 & -0.21 & 0.834 \\
$D_{GFC}$ (2008) & 0.1248 & 0.1354 & 0.92 & 0.357 \\
\bottomrule
\end{tabular}
\end{table}

\subsection{Robustness Checks Summary}

\begin{table}[H]
\centering
\caption{Robustness Checks: Error Correction Term Across Specifications}
\label{tab:robustness_app}
\begin{tabular}{lrrrr}
\toprule
Specification & ECT Coefficient & p-value & R² & N \\
\midrule
R0: Baseline (PCA-12) & -0.5477 & 0.0576 & 0.3824 & 31 \\
R1: Gulf PCA-6 & -0.4815 & 0.0492 & 0.3696 & 31 \\
R2: Trade-Weighted External Demand & -0.4821 & 0.0322 & 0.4312 & 31 \\
R3: Policy Rate + Interest Spread & -1.0217 & 0.0295 & 0.5814 & 31 \\
R4: Excluding Major Shock Years & -0.7264 & 0.0512 & 0.4535 & 28 \\
R5: Post-2000 Subsample & -1.1492 & 0.0000 & 0.8255 & 23 \\
R6: Level Dependent Variable & -0.6583 & 0.0041 & 0.4468 & 31 \\
\bottomrule
\end{tabular}
\end{table}

All replication codes, datasets, and additional outputs are available from the corresponding author upon request.

\end{document}